\documentclass[12pt, letterpaper]{article}

\pdfoutput=1

\usepackage{amsmath}
\usepackage{amsfonts}
\usepackage{amssymb}
\usepackage[colorlinks,citecolor=blue,urlcolor=blue,bookmarks=false,hypertexnames=true]{hyperref}
\usepackage{cite}

\usepackage{graphicx, physics, subcaption}

\numberwithin{equation}{section}

\usepackage{color}

\usepackage{amsmath, amssymb, graphics, epsfig, graphicx}
\usepackage{epsf}
\usepackage{epstopdf}
\usepackage {amssymb}
\newcommand{\nc}{\newcommand}
\nc{\ba}{\begin{eqnarray}}
\nc{\ea}{\end{eqnarray}}

\newcommand{\calR}{{\cal{R}}}
\newcommand{\calP}{{\cal{P}}}

\def\bfk{{\bf k}}

\def\bfx{{\bf x}}

\nc{\cN}{ {\cal{N}} }

\begin{document}

\vspace{5mm}
\vspace{0.5cm}
\begin{center}

\def\thefootnote{\fnsymbol{footnote}}

{\bf\large Stochastic Ultra Slow Roll Inflation }
\\[0.5cm]

{ 
Hassan Firouzjahi$^{1}\footnote{firouz@ipm.ir},$
Amin Nassiri-Rad$^{1}\footnote{amin.nassiriraad@ipm.ir}, $
Mahdiyar Noorbala$^{2,1}\footnote{mnoorbala@ut.ac.ir} $
}
\\[0.5cm]

{\small \textit{$^{1}$ School of Astronomy, Institute for Research in Fundamental Sciences (IPM) \\ P.~O.~Box 19395-5531, Tehran, Iran
}}\\
{\small \textit{$^{2}$ Department of Physics, University of Tehran, Iran, P.~O.~Box 14395-547
}}\\

\end{center}

\vspace{.8cm}

\hrule \vspace{0.3cm}


\begin{abstract}

We study the ultra slow roll model in the context of stochastic inflation. Using stochastic $\delta N$ formalism, we calculate the mean number of $e$-folds, the power spectrum, the bispectrum and the stochastic corrections into these observables.  We reproduce correctly the known  leading classical contributions to these cosmological observables while we show that the fractional corrections to cosmological observables induced from stochastic dynamics  are at the order of power spectrum. In addition, we consider a hypothetical setup containing two absorbing barriers on both sides of the field configuration and calculate the probability of first boundary crossing associated with the classical motion and quantum jumps. This analysis includes the limit of Brownian motion of the quantum fluctuations of  a test scalar field  in a dS spacetime.

\end{abstract}
\vspace{0.5cm} \hrule
\def\thefootnote{\arabic{footnote}}
\setcounter{footnote}{0}
\newpage

\section{Introduction}
\label{sec:intro}

The simplest models of inflation are based on scalar field dynamics in which a scalar field, the inflaton field,   rolls slowly on a flat potential yielding a long period of inflation to solve the flatness and the horizon problems of the standard big bang cosmology.   The quantum fluctuations of the inflaton are the key ingredients of any consistent model of inflation. Indeed, it is believed that the quantum fluctuations of the inflaton field or other light scalar fields during inflation seed the temperature fluctuations and cosmological density perturbations which are observed in CMB maps or in large scale structure.  The basic predictions of models of inflation are that  the primordial perturbations are nearly Gaussian, nearly adiabatic and nearly scale invariant, which are well consistent with cosmological observations \cite{Akrami:2018odb, Ade:2015lrj}.

The stochastic formalism is a powerful method to study quantum fluctuations during 
inflation \cite{Vilenkin:1983xp, Starobinsky:1986fx, Nakao:1988yi, Sasaki:1987gy, Nambu:1987ef, Nambu:1988je,  Kandrup:1988sc,  Nambu:1989uf, Mollerach:1990zf, Linde:1993xx, Starobinsky:1994bd, Kunze:2006tu, Prokopec:2007ak, Prokopec:2008gw, Tsamis:2005hd, Enqvist:2008kt, 
Finelli:2008zg, Finelli:2010sh, Garbrecht:2013coa, Garbrecht:2014dca, Burgess:2014eoa, 
Burgess:2015ajz, Boyanovsky:2015tba,  Boyanovsky:2015jen, Fujita:2017lfu}.  
In this approach, the quantum fluctuations of light scalar fields, such as the inflaton field, are decomposed into the long and short wavelengths perturbations, depending on whether the perturbations are inside the Hubble horizon or outside the Hubble horizon. The small scale perturbations act as the active source of noises for super-horizon scale perturbations. In the simplest models of inflation, these noises are Gaussian with the amplitude $H/2\pi$ in which $H$ is the Hubble expansion rate during inflation.  

The  extension of $\delta N$ formalism to the stochastic setup has been studied in 
\cite{Fujita:2013cna, Fujita:2014tja, Vennin:2015hra} to calculate cosmological correlations \cite{Vennin:2016wnk, Assadullahi:2016gkk, Grain:2017dqa, Noorbala:2018zlv}. The $\delta N$ formalism \cite{Sasaki:1995aw, Sasaki:1998ug, Lyth:2004gb, Wands:2000dp, Lyth:2005fi} is based on the separate Universe approach in which the super-horizon perturbations modify  the background expansion histories of the nearby Universes. It  is proved to be a powerful tool to calculate the curvature perturbation power spectrum, bispectrum and higher order correlations.  In particular, using the stochastic $\delta N$ formalism, Vennin and Starobinsky \cite{Vennin:2015hra} have reproduced the well-known result of Maldacena \cite{Maldacena:2002vr} for the bispectrum of local shape non-Gaussianity   and the consistency condition in single field slow-roll inflationary models. This was revisited recently in  \cite{Abolhasani:2018gyz} in which the kinematical effects of a large scale perturbations on small scale perturbations are calculated to reproduce the Maldacena's consistency condition using the  standard $\delta N$ formalism.  

There were interests to construct examples of single field inflationary models which can violate Maldacena's consistency condition. This has important observational consequence, namely, to what extent one may rule out ``all'' single field inflationary models should Maldacena's consistency condition, relating the  amplitude of local-shape non-Gaussianity $f_{NL}$ to the spectral index $n_{s}-1$,  be violated in cosmological observations. Models of non-attractor inflation are among the very few known examples which can violate the single field non-Gaussianity consistency condition \cite{Namjoo:2012aa, Chen:2013aj, Chen:2013eea}. In the simplest setup of non-attractor inflation, known as the ultra slow-roll (USR) model, the potential is very flat in a finite range of the field value so the kinetic energy falls off exponentially \cite{Kinney:2005vj, Martin:2012pe, Motohashi:2014ppa, Pattison:2018bct}.    In this setup, the would-be decaying mode of curvature perturbation is actually the growing mode, leading to an exponential growth of curvature perturbations on super-horizon scales. This is the key effect which violates the single field non-Gaussianity consistency condition \cite{Bravo:2017wyw, Mooij:2015yka, Akhshik:2015nfa, Akhshik:2015rwa, Finelli:2017fml}. 
Of course, to prevent the arbitrary growth of the curvature perturbation one has to terminate the 
USR phase, say by a waterfall mechanism,  so one has a second (long) stage of attractor inflation in which the curvature perturbation is frozen on super-horizon scales as in conventional models of inflation. The effects of transition from the non-attractor phase to the attractor phase were studied in some details in 
\cite{Cai:2016ngx, Cai:2017bxr}. 

Since in USR models the potential is very flat, then the quantum diffusions associated with the quantum fluctuations of the inflaton field may play important roles. It is an interesting question how one can use the stochastic formalism to study the cosmological perturbations in models of USR inflation. This was studied for the purpose of primordial black hole formation in \cite{Biagetti:2018pjj, Ezquiaga:2018gbw}, see also \cite{Pattison:2017mbe}.  The goal of this paper is to use the stochastic $\delta N$ formalism to calculate the power spectrum and the bispectrum in a simple model of USR inflation. We reproduce the known previous results  and calculate the sub-leading stochastic corrections associated with the quantum diffusion of the inflaton fluctuations. In addition, we calculate the first hitting probabilities of inflaton field  crossing the hypothetical boundaries on either side of the field space due to quantum jumps of the inflaton field. 

The paper is organized as follows. In sections \ref{non-attractor} and \ref{stoc-rev} 
we briefly review the simple model of USR inflation and the method of stochastic inflation which will be used in the follow-up analysis. In section \ref{USR-stoch} we present our analysis of the mean number of $e$-folds, the power spectrum and the bispectrum using stochastic $\delta N$ formalism in USR inflation. In section  \ref{hitting} we present the probabilities of first boundary crossing due to classical motion and quantum jumps. The  
summary and conclusions are  presented in section \ref{summary} while some (important) technicalities associated with higher order correlations of the noise and the bispectrum are relegated to the appendices.

\section{Ultra Slow-Roll  Inflation}
\label{non-attractor}

In this section we briefly review the simplest setup of non-attractor or USR inflation which will be used in our analysis in section \ref{USR-stoch}. 

As in  \cite{Namjoo:2012aa}, we consider a model of inflation with a flat potential $V=V_{0}$ during the first non-attractor phase of inflation. During this phase, the background equations are 
\ba
\label{background}
\ddot \phi +3 H\dot \phi=0\,,
\quad 3 M_{P}^2 H^2 =\dfrac{1}{2} \dot \phi^2 + V_0\simeq V_0\,,
\ea
in which $M_P$ is the reduced Planck mass and $H= \dot a/a$ is the Hubble expansion rate during inflation.
As a result, we have $\dot\phi\propto a^{-3}$ and the first slow-roll parameter $\epsilon \equiv  -\dot H/H^{2}$ falls off exponentially,  $\epsilon \propto a(t)^{-6}$. However,  the second slow-roll parameter $\eta \equiv \frac{\dot \epsilon}{\epsilon H}$ will be very nearly constant and not small, $\eta \simeq -6 + { \cal{O}} (\epsilon^{2})$. 

The evolution of comoving curvature perturbation $\calR$  is given by
\begin{eqnarray}
\left( a^2 \epsilon{\calR}'  \right)' + k^2a^2\epsilon \calR=0\,,
\end{eqnarray}
in which a prime denotes the derivative with respect to the conformal time $\tau$,   $d \tau = dt/H a(t)$.  On super-horizon scales, $k/a H \rightarrow 0$ with $k$ as usual being the Fourier wave number,   the solution is 
\ba
\label{solR}
\calR= C_1+C_2 \int \frac{d\tau}{a^2\epsilon} \, ,
\ea
where $C_1$ and $C_2$ are two constants of integration.
In the  conventional models of single field slow-roll inflation (when the attractor phase has been reached), the term in Eq.~(\ref{solR}) containing $C_2$ represents a decaying mode 
which rapidly falls off on super-horizon scales. However, in the USR setup with the system still being in the non-attractor phase,  $\epsilon \propto a^{-6}$ so the would be decaying mode actually  dominates over the constant mode $C_{1}$. In this limit, we obtain $\calR \propto a^{3}$ so the curvature perturbations grow exponentially on super-horizon scale. 

The power spectrum of curvature perturbations and the local type non-Gaussianity can be calculated using either the field theoretic in-in approach or the $\delta N$ formalism. In the latter approach, we have to express the number of $e$-folds $N$ as a function of the background quantities $\phi$ and $\dot \phi$. Note that since the system has not reached the attractor phase, then $\phi$ is not a clock so we have to solve $N$ as a function of both $\phi$ and $\dot \phi$, i.e. $N= N(\phi, \dot \phi)$. This is the key difference of the $\delta N$ approach in USR model compared to conventional models  in which the system has reached the attractor phase and $N= N(\phi)$.

Solving the background field equations (\ref{background}), we obtain
\ba
N(\phi,\dot\phi)=\frac{1}{3}\ln
\Big[\frac{\dot\phi}{\dot\phi+3H(\phi-\phi_e)}\Big]\,, 
\label{Nform}
\ea
in which $\phi_{e}$ is the value of $\phi$ at the end of USR phase.
Eq.~(\ref{Nform}) gives $N$ as a function of $(\phi, \dot \phi)$ in phase space.  Note that the convention used in  \cite{Namjoo:2012aa} is such that $d N = - H dt$ with the number of $e$-folds counted backward from the surface of end of inflation, so $N\ge 0$, with $N(\phi_{e}, \dot \phi_{e})=0$. 

Note the curious effect that there is a limit $\phi= \phi_{\mathrm {max}}$
beyond which the field $\phi$ cannot go further classically. This is because $\dot \phi $ falls off exponentially so  if inflation is not turned off (say via a waterfall field mechanism) then it takes $\phi$ an infinite time to reach $\phi_{\mathrm {max}}$. Denoting  the initial values of $\phi$ and its velocity by $\phi_0$ and $\dot \phi_0$ respectively, then $N$ in Eq.~(\ref{Nform}) diverges for $\phi \rightarrow \phi_{\mathrm {max}}$ in which 
\ba
\label{phimax}
\phi_{\mathrm {max}} = \phi_0 + \frac{\dot \phi_0}{3 H} \, .
\ea
To have a finite period of USR inflation, we require $\phi_e < \phi_{\mathrm {max}}$.

To use $\delta N$ formalism, we have to find the amplitude of $\delta \phi_{k}$ fluctuations on the initial flat slicing which is obtained to be 
\begin{eqnarray}
\label{phi-k}
\phi_k=\frac{H}{(2k)^{3/2}}(1+ik\tau)e^{-ik\tau}\, .
\end{eqnarray}
Note that, as in conventional models of inflation, $\delta \phi$ freezes on super-horizon scales,  $ \delta \dot \phi \simeq 0\,  (k/a H \ll1)$. As a result, although we have to keep track of  $N$ as a function of $\dot \phi$ at the background level, we can neglect the contribution of $ \delta \dot \phi$ when perturbing $N$. 

Now, using the $\delta N$ formalism to second order in perturbations, and remembering that  $\delta \dot \phi \simeq 0 $, we obtain
\ba
\label{deltaN-USR}
\delta N
&\simeq&\frac{\partial N}{\partial\phi}\delta\phi
+\frac{1}{2}\frac{\partial^2 N}{\partial\phi^2}\delta\phi^2 \nonumber\\
&=&\frac{-H}{\dot\phi+3H(\phi-\phi_e)}\delta\phi  
+\frac{3H^2}{2\Bigl(\dot\phi+3H(\phi-\phi_e)\Bigr)^2}\delta\phi^2 \, .
\ea

From Eqs.~(\ref{phi-k}) and (\ref{deltaN-USR}),  the power spectrum of curvature perturbation $\calR = \delta N$,  calculated at the end of USR phase where $\phi=\phi_{e}$,  is  obtained to be
\ba
\label{Pe}
{\cal P}_e \equiv \dfrac{k^3}{2 \pi^2} P_k\simeq
\dfrac{H^2}{8 \pi^2 M_{P}^2 \epsilon_e}  \, ,
\ea
in which $\epsilon_{e}$ is the value of $\epsilon$ at the end of USR phase.  

Finally, from Eq.~(\ref{deltaN-USR}), the non-Gaussianity parameter $f_{NL}$, as defined in Eq.~(\ref{fNL-def}),   is easily obtained to be $f_{NL} = \frac{5}{2}$. 

As mentioned before, this model by itself is not consistent. A rapid growth of $\calR$ will make the setup non-perturbative. One requires a mechanism, such as a sudden waterfall instability, to terminate the USR stage so a second long slow-roll phase is followed after the initial short USR phase.

\section{Review of Stochastic Inflation}
\label{stoc-rev}

Here we briefly review the formalism of stochastic inflation which will be used in our analysis in next sections. Here we will mainly follow \cite{Nakao:1988yi, Sasaki:1987gy}. 

Considering a single field model of inflation with the potential $V(\phi)$,  the Klein-Gordon equation   is given by 
\begin{equation}
\label{2}
    \left(\frac{\partial^2}{\partial t^2}+3H\frac{\partial}{\partial t}-\frac{\nabla^{2 } }{a^{2}} \right)\phi\left(\bfx,t\right)+\frac{\partial V }{\partial\phi}(\bfx, t)=0.
\end{equation}
Note that in obtaining the above equation we have neglected the gravitational back-reactions, i.e. we did not perturb the metric. The motivation is that we would like to use the stochastic $\delta N$ formalism in which the initial conditions of the scalar field quantum perturbations are calculated on the spatially flat hypersurfaces, so upon appropriate choice of gauge, we can neglect the spatial metric perturbations. Furthermore, the shift and the lapse functions in the ADM decompositions are higher order in gradient expansions of the separate Universe approach \cite{Lyth:2004gb} so one can neglect their contributions as well in the $\delta N$ formalism.

We split $\phi$ and its conjugate momentum $v$ into the short and long wavelengths as follows 
\begin{equation}
\label{3}
\begin{split}
\phi\left(\bfx,t\right)=\phi_l\left(\bfx,t\right)+\sqrt{\hbar}\phi_s\left(\bfx,t\right),
\end{split}
\end{equation}
\begin{equation}
\label{4}
\begin{split}
v\left(\bfx,t\right)=v_l\left(\bfx,t\right)+\sqrt{\hbar}v_s\left(\bfx,t\right),
\end{split}
\end{equation}
where $l$ and $s$ denote the long modes and short modes respectively. Furthermore,   the short modes satisfy the following decomposition in Fourier space 
\begin{equation}
\label{5}
    \phi_s\left(x,t\right)=\int\frac{d^3k}{\left(2\pi\right)^3}\theta\left(k-\varepsilon aH\right)\phi_\bfk\left(t\right)e^{ik.x},
\end{equation}
\begin{equation}
\label{6}
    v_s\left(x,t\right)=\int\frac{d^3k}{\left(2\pi\right)^3}\theta\left(k-\varepsilon aH\right)\dot\phi_\bfk\left(t\right)e^{ik.x}.
\end{equation}
Here $ \varepsilon$ is a small dimensionless number $\varepsilon \ll 1$ which is introduced to separate the large and small scales in an appropriate way. 
The factor $\sqrt \hbar$ has been inserted for the short modes in Eqs.~(\ref{3}) and (\ref{4})  to specify the quantum natures of the short modes. In addition, the operator
$\phi_{\bfk }(t)$ satisfies  $\phi_\bfk=a_\bfk\varphi_k+a^\dagger_{-\bfk}\varphi_{-k}^*$ and $\varphi_k$ is the positive frequency mode function satisfying the the Klein-Gordon equation.

 By expanding Eq.~\eqref{2} around $\phi_l$ and $v_l$ up to first order of $\sqrt{\hbar}$ we get the following equations of motion for $\phi_l$ and $v_l$ \cite{Nakao:1988yi, Sasaki:1987gy}
\ba
    \label{7}
    \dot \varphi_l &=&v_l+\sqrt{\hbar}\sigma,\\
    \label{8}
    \dot{v}_l &=&-3Hv_l+\frac{1}{a^2}\nabla^2\varphi_l-V'\left(\varphi\right)+\sqrt{\hbar}\tau,
\ea
where $\sigma$ and $\tau$ are given by
\begin{equation}
    \label{9}
    \sigma\left(\bfx,t\right)= \varepsilon aH^2\int\frac{d^3 \bfk}{\left(2\pi\right)^3}\delta\left(k-  \varepsilon aH\right)\phi_\bfk\left(t\right)e^{i \bfk \cdot \bfx},
\end{equation}
\begin{equation}
    \label{10}
    \tau\left(\bfx,t\right)= \varepsilon aH^2\int\frac{d^3 \bfk}{\left(2\pi\right)^3}\delta\left(k-  \varepsilon aH\right)\dot\phi_\bfk\left(t\right)e^{i \bfk \cdot \bfx}.
\end{equation}
Note that the short modes $\phi_{s}$ and $v_{s}$ play the roles of the source terms for the evolution of the long modes $\phi_{l}$ and $v_{l}$ via  $\tau$ and $\sigma$ which appear in  the right hand side of Eqs.~(\ref{7}) and (\ref{8}). 

Starting with the Bunch-Davies initial condition $|0 \rangle$, the correlation function of these sources are given by \cite{Nakao:1988yi, Sasaki:1987gy}
\ba
    \label{11}
   & \left \langle 0\left|\sigma\left(\bfx_1\right)\sigma\left(\bfx_2\right)\right| 0\right\rangle 
    \approx \varepsilon^{\frac{2M^2}{3H^2}}\frac{H^3}{4\pi^2}j_0\big(\varepsilon aH | \bfx_1- \bfx_2 |\big)\delta\left(t_1-t_2\right),\\
    \label{12}
   & \left \langle 0\left|\tau\left( \bfx_1\right)\tau\left( \bfx_2\right)\right| 0\right\rangle
    \approx \varepsilon^{\frac{2M^2}{3H^2}}\big(\frac{M^2}{3H^2}+\varepsilon^2\big)^2\frac{H^5}{4\pi^2}j_0\big(\varepsilon aH | \bfx_1-\bfx_2 |\big)\delta\left(t_1-t_2\right),\\
    \label{13}
    & \langle 0\left|\sigma( \bfx_1)\tau( \bfx_2)+\tau( \bfx_2)\sigma( \bfx_1) | 0\right\rangle  
    \approx -2 \varepsilon^{\frac{2M^2}{3H^2}} 
    \big(\frac{M^2}{3H^2}+ \varepsilon^2\big)\frac{H^4}{4\pi^2}j_0\big( \varepsilon aH | \bfx_1-\bfx_2|\big)\delta (t_1-t_2) 
\ea
where $M^2$ is the average mass of the long wavelength component of the field and $j_0$ is the zeroth order Bessel function. 

In addition,  the correlation function of $\sigma$ and $\tau$ are given by 
\ba
    \label{correlation}
    \left[\sigma\left(\bfx_1\right),\sigma\left(\bfx_2\right)\right] &=&\left[\tau\left(\bfx_1\right),\tau\left(\bfx_2\right)\right]=0, \\
    \label{correlation2}
    \left[\sigma\left( \bfx_1\right),\tau\left(\bfx_2\right)\right] &= & i \varepsilon^3\frac{H^4}{4\pi^2}j_0\big( \varepsilon aH | \bfx_1-\bfx_2|\big)\delta\left(t_1-t_2\right).
\ea
As it can be seen from Eqs.~\eqref{correlation} and  \eqref{correlation2},  the quantum nature of $\sigma$ and $\tau$ disappears if we choose $\varepsilon$ small enough. 

The above formalism was general, without specifying the form of the potential. Now consider our USR case in which  $V\left(\phi\right)=V_0$, then clearly $M^2=0$, and from Eqs.~\eqref{11}-\eqref{13} we see that the $\varepsilon$ dependence of $\sigma$ disappears and $\tau=O\left( \varepsilon^2\right)$.  So if $ \varepsilon$ is chosen small enough we can neglect $\tau$ and write
\begin{equation}
    \label{stoch}
     \left<0\left|\sigma\left(\bfx_1\right)\sigma\left(\bfx_2\right)\right|0\right>\approx\frac{H^3}{4\pi^2}\delta\left(t_1-t_2\right)=\frac{H^4}{4\pi^2}\delta\left(N_1-N_2\right).
\end{equation}
In the second equality we have changed the time variable to the number of $e$-folds via $N=Ht$. Note that in the above limit  $\sigma$ is only time dependent. Hence, in the super horizon limit, i.e $k \ll aH$, and dropping  the subscript $l$  for convenience,  one can write Eqs.~\eqref{7} and  \eqref{8} for the coarse grained long modes as follows
\ba
    \label{17}
        \frac{d\phi}{dN} &=&\frac{v}{H}+\frac{H}{2\pi}\xi\left(N\right), \\
    \label{18}
        \frac{dv}{dN}&=&-3v,
\ea
where we have set $\sigma\equiv\frac{H}{2\pi}\xi$ so $\xi$ is a normalized white classical noise satisfying 
\ba
\label{xi-noise}
\big \langle \xi\left(N\right)\big \rangle = 0 \, , \quad \quad 
\big \langle \xi\left(N\right)\xi\left(N'\right)\big \rangle =\delta\left(N-N'\right) \, .
\ea 

Note the curious conclusion that while $\phi$ satisfies a stochastic differential equation with the noise $\xi (N)$, but  the evolution of $v$ is deterministic. This is a consequence of the conclusion that $\tau=O\left( \varepsilon^2\right)$. 

One can easily solve Eq.~\eqref{18}, obtaining 
\begin{equation}
    \label{19}
    v\left(N\right)=\dot\phi_0e^{-3N} \, ,
\end{equation}
in which $\dot\phi_0$ is a constant of integration, corresponding to the initial velocity at the start of USR phase where we have set $N=0$. 

Substituting Eq.~\eqref{19} into Eq.~\eqref{17} yields the following Langevin equation
\begin{equation}
    \label{Langevin}
    \frac{d\phi}{dN}=\frac{\dot\phi_0}{H}e^{-3N}+\frac{H}{2\pi}\xi\left(N\right).
\end{equation}
By the initial condition $\phi\left(0\right)=\phi_0$, Eq.~\eqref{Langevin} can be integrated to yield 
\begin{equation}
    \label{21}
    \phi\left(N\right)= \phi_0 + \frac{\dot\phi_0}{3H}\left(1-e^{-3N}\right)+\frac{H}{2\pi}W\left(N\right),
\end{equation}
where 
\ba
\label{WN}
W\left(N\right)\equiv\int^N_0\xi\left(N\right)dN \, ,
\ea  
is the  Wiener process associated with the noise $\xi(N)$ \cite{Evans}.

To obtain Eq.~(\ref{21}) we have assumed that $H$ is very nearly constant so we have neglected its 
evolution during the USR phase. This is well justified, since during the USR phase $\epsilon$ falls off 
like $a^{-6}$ so to leading order in $\epsilon$ one can safely neglect the evolution of $H$. Equation (\ref{21}) is the key equation for our follow up analysis in next sections. 

\section{Stochastic Analysis of USR Inflation}
\label{USR-stoch}

In this section we present our analysis of the stochastic corrections into various cosmological correlations, such as the mean number of $e$-folds, the power spectrum and bispectrum. As we discussed previously, in the USR setup the potential is very flat in some ranges of the field displacement. Therefore, during this period,  the spacetime is very close to a dS spacetime and one expects that the stochastic quantum jumps of the inflaton field play important roles in the evolution of its trajectory.  The goal of this analysis is to calculate the leading stochastic corrections in cosmological correlations. 

In our specific USR setup described in section \ref{non-attractor}, the surface of end of inflation is determined by $\phi=\phi_{e}$. In addition, the initial values of the field and its velocity in phase space are also given quantities, defined 
by $\phi_{0}$ and $\dot \phi_{0}$ respectively.  However, note that the total number of $e$-folds starting from the initial point $(\phi_{0}, \dot \phi_{0})$ in phase space to the final point $\phi = \phi_{e}$ is a stochastic variable. The reason is as follows. Because of the quantum kicks of the inflaton field, the trajectory of the field and its velocity will be a random process, very much similar to a Brownian process. There are infinite trajectories in phase space for the system to start from  $(\phi_{0}, \dot \phi_{0})$ and to end at $\phi=\phi_{e}$. Each path in phase space represents one particular realization of inflation so in this view the total number of $e$-folds is a stochastic process. We denote the total number of $e$-folds in each realization by $\cN$ in order to distinguish it from $N$, which is the usual clock. 


With these discussions in mind, the quantities of interests are the mean number of $e$-folds
$\big \langle \cN \big \rangle $, the power spectrum $\calP_{\calR}$ and the amplitude of bispectrum $f_{NL}$.  While the calculation of $\big \langle \cN \big \rangle $ is direct, we need some dictionaries of stochastic $\delta N$ formalism to calculate the power spectrum and bispectrum. 

Similar to the logic of \cite{Fujita:2013cna, Fujita:2014tja, Vennin:2015hra},  starting with the $\delta N$ formula, $\calR= \delta \cN$, let us look at the variance of curvature perturbation at each point $\bfx$, $\langle \calR^{2}(\bfx) \rangle$:
\ba
\label{variance-R}
\langle \calR^{2}(\bfx)  \rangle = \int_{k_{i}  }^{k_{e}} \frac{dk }{k} 
\frac{k^{3}}{2 \pi^{2}} | \calR_{k}|^{2} \simeq  
\int_{\ln k_{e} - \langle \cN \rangle }^{\ln k_{e}} d N\, 
\calP_{\delta N} \, ,
\ea
in which $\calP_{\calR} = (k^{3}/2 \pi^{2})  \big|\calR_{k}\big|^{2}$ is the dimensionless power spectrum and $k_{i}$ and $k_{e}$ respectively are the first and the last modes which leave the Hubble radius during USR phase of inflation.  Note that up to sub-leading slow-roll $\epsilon$ corrections, $\langle N \rangle \simeq \ln(k_{i}/k_{e})$ which was used to change the domain of integration in  second integral. Note that because of the background translation invariance, the variance is independent of the choice of $\bfx$.

The first integral in Eq.~(\ref{variance-R}) represents the accumulative effects of the modes which have left the horizon, from the start  of the USR phase to any given intermediate time, and will modify the background FRW expansion for all modes which are still sub-horizon.  This is the spirit of the separate Universe approach in which the effect of a long mode is to modify the background expansions of the nearby FRW patches. 

On the other hand, from the definition of the variance of $ \cN$ as an stochastic variable,  we have 
\ba
\label{deltaN-def}
\langle \delta \cN^{2}  \rangle  \equiv \big \langle \big(\, \cN - \langle \cN \rangle \, \big)^{2}  \big\rangle
= \langle \cN^{2 } \rangle - \langle \cN \rangle^{2} \, .
\ea
Now, combining Eqs.~(\ref{variance-R}) and (\ref{deltaN-def}) we can relate $\calP_{\calR}$ to the derivative of $\delta \cN$ as follows
\ba
\label{power}
    \mathcal{P}_\calR= \mathcal{P}_{\delta\mathcal{N}}=\frac{\mathrm d\left<\delta\mathcal{N}^2\right>}{\mathrm d\left<\mathcal{N}\right>}.
\ea

Using the same strategy, the bispectrum is related to $\langle \calR^{3}(\bfx)  \rangle$
which can be used to calculate the amplitude of local non-Gaussianity $f_{NL}$. We present the details of the corresponding analysis in the Appendix \ref{fNL-app} where it is shown that $f_{NL}$ is related to the second derivative of the third moments of $ \cN$ as follows 
\ba
\label{fnl}
  f_{NL}=\frac{5}{36 \calP_{\calR}^{2}}\frac{\mathrm d^2\left<\delta\mathcal{N}^3\right>}{\mathrm d \left\langle \mathcal{N}\right\rangle^2} \, ,
\ea
in which $\langle \delta \cN^{3} \rangle \equiv \big \langle \big(\, \cN - \langle \cN \rangle \, \big)^{3}  \big\rangle $.


\subsection{Mean Number of $e$-folds}
Here we calculate the mean of the total number of $e$-folds $\left\langle \mathcal{N}\right\rangle$. 

The evolution of $\phi(N)$ is obtained in Eq.~\eqref{21}. We can rewrite Eq.~\eqref{21} 
to obtain $\cN$  where  now its is understood that $N= \cN$ is the total number of $e$-folds during the USR phase.  We remind the reader that  we use the convention that at the start of  USR phase $\cN=0$. 

It is convenient to define $N_{c}$ as the total number of $e$-folds in the classical limit, i.e., in the absence of stochastic kicks. From the analysis of section (\ref{non-attractor}), Eq.~(\ref{Nform}), we have 
\ba
\label{Nc-def}
N_{c}= -\frac{1}{3} \ln \Big[  1-\frac{3H}{\dot\phi_0}(\phi_{e} - \phi_{0})\Big] \, .
\ea
Now, solving Eq.~\eqref{21}  for $\cN$ we have
\ba
\label{N-eq1}
e^{-3 \cN} = e^{-3 N_{c}} \big (1+ \kappa W(\cN) \big)  \, ,
\ea
in which we have defined the parameter $\kappa$ via
\ba
\kappa \equiv  3\,  e^{3 N_{c}} \Big( \frac{H^{2}}{2 \pi \dot \phi_{0}} \Big)  
\equiv 3\,  e^{3 N_{c}} \sqrt{\calP_{0} } = 3  \sqrt{\calP_{e} } \, ,
\ea
where in the second equation we have defined $\calP_{0}$ as the  power spectrum at the start of the USR phase in the absence of stochastic effects. Similarly, $\calP_{e}$ is defined as the  power spectrum at the end of the USR phase in the absence of stochastic effects, given in Eq.~(\ref{Pe}).  Note that, in the absence of stochastic effects, the curvature perturbations on super-horizon grows like $\calR \propto a^{-3}$ so we have $\calP_{e} = e^{6 N_{c}} \calP_{0}$. 

Assuming that the system is perturbative, i.e., the cosmological perturbations are small, we require from the observations that $\calP_{e} \sim 10^{-9}$ so $\kappa \ll1$. As we will see, the parameter $\kappa $ is the expansion parameter of our stochastic analysis. This makes sense.   In one $e$-fold, the quantum jump of the inflaton field is $H/2 \pi$, while its classical roll is of order $e^{-3N_c}\dot\phi_0/H$.  The ratio of these two is thus encoded  in the parameter $\kappa$. 

Solving Eq.~(\ref{N-eq1}) perturbatively as a series of $\kappa$, we have 
\ba
    \label{5-7}
\mathcal{N}=N_{c}+\frac{1}{3}\sum_{n=1}^{\infty}\frac{\left(-1\right)^n}{n}\kappa^nW^n\left(\mathcal{N}\right) \, .
\ea
Now taking the stochastic average of both sides we obtain
\ba
    \label{average}
    \big\langle \mathcal{N} \big\rangle = N_{c}+\frac{1}{3}\sum_{n=1}^{\infty}\frac{\left(-1\right)^n}{n}\kappa^n 
    \big \langle W^n\left(\mathcal{N}\right)  \big\rangle\, .
\ea
To proceed further, we need to calculate $\big \langle W(\cN)^{n} \big\rangle$ for $n=0, 1, 2,... .$.
This analysis is  non-trivial as $W(\cN)$ depends  on $\cN$ which itself is a stochastic variable. It turns out that to calculate the first stochastic corrections to $f_{NL}$ we need to calculate $\big \langle W(\cN)^{n} \big\rangle$ up to $n=6$.  We have presented the corresponding analysis in Appendix \ref{calculus}, where using the Ito calculus of stochastic processes, we have shown that 
\ba
\label{W12}
\big\langle W(\cN) \big\rangle =0 \, ,  \quad
\big \langle W(\cN)^{2} \big\rangle = \big \langle \cN  \big\rangle  \, .
\ea
\ba
\label{W3}
\big\langle W(\cN)^{3}\big \rangle =  - \kappa N_c \big[ 1+ \kappa^2 (N_c + \frac{2}{3} )   \big] + {\cal O}(\kappa^5)  \, ,
\ea
\ba
\label{W4}
\big\langle W(\cN)^{4} \big\rangle = 3 N_c^2 + \kappa^2 \big( 3 N_c^2 + \frac{5}{3} N_c \big)
+ {\cal O}(\kappa^{4})  \, .
\ea
\ba
\label{W5}
 \big\langle W(\cN)^{5}  \big\rangle = -10 \kappa N_c^2 + {\cal O}(\kappa^{3}) 
\ea
and
\ba
\label{W6}
 \big\langle W(\cN)^{6}  \big\rangle = 15 N_c^3  + {\cal O}(\kappa^{2})  \, .
\ea

Plugging the above formulas in Eq.~(\ref{average}) we obtain 
\ba
    \label{average2}
    \big\langle \mathcal{N} \big\rangle =N_{c} \Big[ 1+   \frac{\kappa^{2}}{6} +  \frac{\kappa^{4}}{36} \big(5 + 9 N_c \big)
    +  \frac{\kappa^{6}}{72} \big(17 + 77 N_c + 60 N_c^2 \big)  \Big]+O\left(\kappa^7\right).
\ea
As expected, the leading term is given by $N_{c}$ the classical value of the number of $e$-folds in the absence of stochastic kicks, as calculated in \cite{Namjoo:2012aa}. The stochastic corrections start at the order $\kappa^{2} \sim \calP_e$. As argued before, this makes sense since the quantum jumps of the inflaton field, once translated in terms of  the curvature perturbations, are naturally encoded in the parameter $\kappa$.

\subsection{Power Spectrum}
As discussed earlier to calculate the power spectrum we need to calculate $\left<\delta\mathcal{N}^2\right> =  \langle \big(\mathcal{N}-\langle\mathcal{N} \rangle \big)^2  \rangle$. Starting with 
the general form of $\cN$ given in Eq.~\eqref{5-7}, we have
\ba
\label{deltaN2-1}
\langle\delta\mathcal{N}^2\rangle &=&
  \frac{\kappa^2}{9} \big\langle W(\cN) ^2 \big\rangle 
 - \frac{\kappa^3}{9} \big\langle W(\cN) ^3 \big \rangle  +  \frac{\kappa^4}{108} \big(   {11 \big\langle W(\cN) ^4 \big\rangle}  - 3 {\big\langle W(\cN)^2 \big\rangle ^2  } \big)  \nonumber\\
  &+& \frac{\kappa^5}{54} \big (  - 5 \big\langle W(\cN) ^5 \big\rangle + 2  \big\langle W(\cN) ^2 \big\rangle \big\langle W(\cN) ^3 \big\rangle
  \big)\nonumber\\
  &+&  \frac{\kappa^6}{1620} \big (  137  \big \langle W(\cN) ^6  \big\rangle -45   \big\langle W(\cN) ^2 \big\rangle \big\langle W(\cN) ^4 \big\rangle
  -20  \big \langle W(\cN) ^3 \big \rangle^2 
  \big) + {\cal O}(\kappa^{7}) .
\ea
Using the expressions found for $ \langle W(\cN)^{3} \rangle, \langle W(\cN)^{4} \rangle $ and $ \langle W(\cN)^{5} \rangle $ from  Eqs.~(\ref{W3}), (\ref{W4}) and (\ref{W5}), we find 
\ba
 \left\langle \delta\mathcal{N}^2\right\rangle = \frac{\kappa^{2}}{9} N_c \Big[ 1+ \frac{\kappa^{2}}{6} (7+ 15 N_c) + \frac{\kappa^{4}}{12}  \big(28+ 143 N_c  + 128 N_c^2 \big)
 \Big]  \,  + {\cal O}(\kappa^{7}) \, .
\ea
The power spectrum is determined by  Eq.~(\ref{power}). Since $\left<\delta\mathcal{N}^2\right>$ and 
$\langle \cN \rangle$ are functions of $N_c$, we can use the chain rule of derivatives to obtain 
\ba
\label{power-final}
\calP_{\calR} &=& \frac{\mathrm d\left<\delta\mathcal{N}^2\right>}{\mathrm d\left<\mathcal{N}\right>}
= \frac{ \left<\delta\mathcal{N}^2\right>'}{ \langle \cN \rangle '} \nonumber \nonumber\\
&=&
\frac{\kappa^{2}}{9}   \Big( 1+ \kappa^{2} (1+ 5 N_c) \Big)  = \calP_{e} \Big(1 + 9 \calP_{e} (1+ 5 N_c) \Big)
+ {\cal O}(\kappa^{6}) \, ,
\ea
in which the prime indicates the derivative with respect to $N_c$.

This is an interesting result. The leading term is $\calP_{e}$, defined in Eq.~(\ref{Pe}), which is in agreement with \cite{Namjoo:2012aa} while the stochastic corrections are at the order $\calP_{e}^{2}$. There are  corrections of higher orders of $\calP_{e}$ which we have discarded in Eq.~(\ref{power-final}).  As a result, we see that the stochastic effects cannot enhance the curvature perturbations to amplify the initial amplitude for the primordial black hole formation.

\subsection{Bispectrum}
Now we calculate the non-Gaussianity parameter $f_{NL}$. For this purpose, we need to calculate 
$\delta \cN^{3} =  \langle \big( \cN - \langle \cN \rangle \big)^{3} \rangle $.  Following the same steps as 
in the case of $\delta \cN^{2}$ we have 
\ba
\label{deltaN3-1}
\big\langle\delta\mathcal{N}^3 \big\rangle &= &
  -\frac{\kappa^3}{27} \big\langle W(\cN) ^3 \big\rangle  +  \frac{\kappa^4}{18} \Big[  \big\langle W(\cN) ^4 \big\rangle
  -  \big\langle W(\cN) ^2 \big\rangle ^2\Big] \nonumber\\
 &+&  \frac{\kappa^5}{108} \Big[ - {7}  \big\langle W(\cN) ^5 \big\rangle
 +  {10}   \big\langle W(\cN) ^2\big \rangle \big\langle W(\cN) ^3 \big\rangle    \Big]   \\
 &+&  \frac{\kappa^6}{216} \Big[  {15}  \big\langle W(\cN) ^6 \big\rangle
 -  {17}   \big\langle W(\cN) ^4\big \rangle \big\langle W(\cN) ^2 \big\rangle   +  2{ \big\langle W(\cN) ^2 \big\rangle^3}   
  -8{ \big\langle W(\cN) ^3 \big\rangle^2}  
  \Big]  + {\cal O} (\kappa^{7})\nonumber \, .
  \ea
Using the formulas for $ \langle W(\cN)^{3} \rangle, \langle W(\cN)^{4} \rangle,  \langle W(\cN)^{5} \rangle$ and $ \langle W(\cN)^{6} \rangle $ from  Eqs.~(\ref{W3}), (\ref{W4}), (\ref{W5}) and (\ref{W6}), and after a long but otherwise straightforward calculation, we obtain
\ba
 \big\langle \delta \cN^{3}  \big\rangle = \frac{\kappa^4}{27} \Big( {N_c} + 3{N_c^2} \Big)
+\frac{ \kappa^6 }{162}\Big( {19 N_c} +  {120 N_c^2} +  {132 N_c^3} 
\Big)
+ {\cal O} (\kappa^{8}) \, .
\ea
Correspondingly, $f_{NL}$ from Eq.~(\ref{fnl}), after some chain derivatives with respect to $N_c$,  is obtained to be
\ba
  f_{NL}=\frac{5}{36 \calP_{\calR}^{2}}\frac{\mathrm d^2\left\langle \delta\mathcal{N}^3\right \rangle }{\mathrm d \left\langle \mathcal{N}\right\rangle^2}  
  &=& \frac{5}{36(\langle \delta \cN^{2} \rangle')^2 } \Big[ {\langle \delta \cN^{3} \rangle''}
  - {\langle \delta \cN^{3} \rangle'} \frac{\langle \cN \rangle''}{\langle \cN \rangle'}
  \Big] \nonumber\\
&=& \frac{5}{2} + \kappa^2 \left(\frac{65}{6} + 30N_c  \right)  + {\cal O} (\kappa^{4}) \, .
\ea
The leading term for $f_{NL}$ agrees exactly with the result of \cite{Namjoo:2012aa} while we also have the sub-leading stochastic corrections  in $f_{NL}$ at the order of $\calP_e$. The stochastic effects induce sub-leading quantum corrections into Maldacena's consistency condition.

\section{Boundary Crossing Probabilities}
\label{hitting}

As an application of stochastic formalism here we consider a hypothetical setup in which we have two 
absorbing barriers in field space located at $\phi_+ > \phi_0$ and $\phi_- <\phi_0$. We assume that inflation ends when the field hits either of the barriers. We would like  to calculate  the first boundary crossing probabilities $p_+$ and $p_-$ which are  the probabilities of hitting first either $\phi_+$ or $\phi_-$ respectively. Note that  it may take a large number of $e$-folds for the field to hit either barrier so this question is not directly relevant for the observable  inflationary period.  Note that since the total probability of hitting either barrier is unity, i.e., we wait long enough that  the field hits either barrier for sure, we have $p_+ + p_- =1$.   

The starting equation is (\ref{21}) which solves $\phi$  as a function of $\cN$ in which now $\cN$ is defined as the total number of $e$-folds required for the field $\phi$ to hit either of the barriers. Without loss of generality we assume  $\phi_0=0$, which is allowed because the potential is shift symmetric,  and $\dot \phi_0 \ge 0$.  Then from Eq.~(\ref{21}) we have 
\ba
\label{phi-cN}
\phi \left(\cN \right)=\frac{\dot\phi_0}{3H}\left(1-e^{-3 \cN}\right)+\frac{H}{2\pi}W\left(\cN\right) \, .
\ea   
In general it is not easy to solve the above equation, involving two absorbing barriers at  $\phi_+$ and $\phi_-$, analytically and one may  require  numerical analysis. However, we can solve the above equation in some interesting limits as we consider below.  

\subsection{Brownian limit}

An interesting limit is when the field has no classical velocity, $\dot \phi_0=0$, so the classical drift term in Eq.~(\ref{phi-cN}) vanishes and
\ba
\label{Brownian-eq}
\phi \left(\cN \right)=\frac{H}{2\pi}W\left(\cN\right) \, .
\ea
This corresponds to a pure Brownian limit in which the field evolves under quantum kicks with the amplitude 
$H/2\pi$ as given by the noise term $W(\cN)$. Of course, this limit is not realistic for the purpose of inflation as the field  $\phi$ is a test field and  has no classical evolution so there is no notion of curvature perturbations. However, this limit is insightful to understand the  
stochastic  effects in dS backgrounds. 

Taking the expectation of Eq.~(\ref{phi-cN}) we obtain 
\ba
\label{Brownian-phi}
\langle \phi(\cN) \rangle =0 \, .
\ea 
On the other hand, from the definition of $p_+$ and $p_-$ we have 
\ba
\label{phi-av-Brownian}
\langle \phi(\cN) \rangle = p_+ \phi_+ + p_- \phi_- \, .
\ea
Combining this with Eq.~(\ref{Brownian-phi}) we  obtain
\ba
\label{Brownian-probs}
p_+ = \frac{-\phi_-}{\phi_+ - \phi_-} \, , \quad \quad
p_- = \frac{\phi_+}{\phi_+ - \phi_-} \, .
\ea
We see that the first hitting probability for a given barrier  is proportional to the distance of the mirror barrier to the origin. The further away the mirror barrier, the higher the probability to  first hit the given barrier.  
In the limit that $\phi_+$  ($\phi_-$) is pushed to infinity, then $p_+ (p_-)$ vanishes which is consistent with  intuition. 

To obtain $\langle \cN \rangle$, we take the expectation value of the square of Eq.~(\ref{phi-cN}), yielding 
\ba
\big \langle \phi(\cN)^2 \rangle =  \big( \frac{H}{2\pi} \big)^2 \langle \cN \rangle \, .
\ea
On the other hand
\ba
\label{phi2-av-Brownian}
\langle \phi(\cN)^2 \rangle = p_+ \phi_+^2 + p_- \phi_-^2 \, .
\ea
Using the values of $p_\pm$ obtained in Eq.~(\ref{Brownian-probs}), we obtain
\ba
\label{N-av-Brownian}
 \langle \cN \rangle =  \frac{-\phi_- \phi_+}{ \big( \frac{H}{2\pi} \big)^2} =
 \Big(  \frac{\phi_+}{\frac{H}{2\pi}} \Big)  \Big( \frac{-\phi_-}{\frac{H}{2\pi}}  \Big) \, .
\ea
To interpret the above result, note that $H/2\pi$ represents the length of each quantum jump so the ratios 
$\phi_+/(H/2\pi)$ and  $-\phi_-/(H/2\pi)$ respectively measure the classical displacements of $\phi_+$ and $-\phi_-$ relative to  quantum jumps to reach the two barriers. 

Note that if the initial position of the field is located on the position of a   barrier then we obtain the expected result that $\langle \cN \rangle =0$ and one of
$p_\pm$ is equal to unity while the other one is zero. For example, if we have $\phi_+=0$, then $p_+=1$ and $p_-=0$. 

\begin{figure}
		\centering
		\includegraphics[scale=0.32]{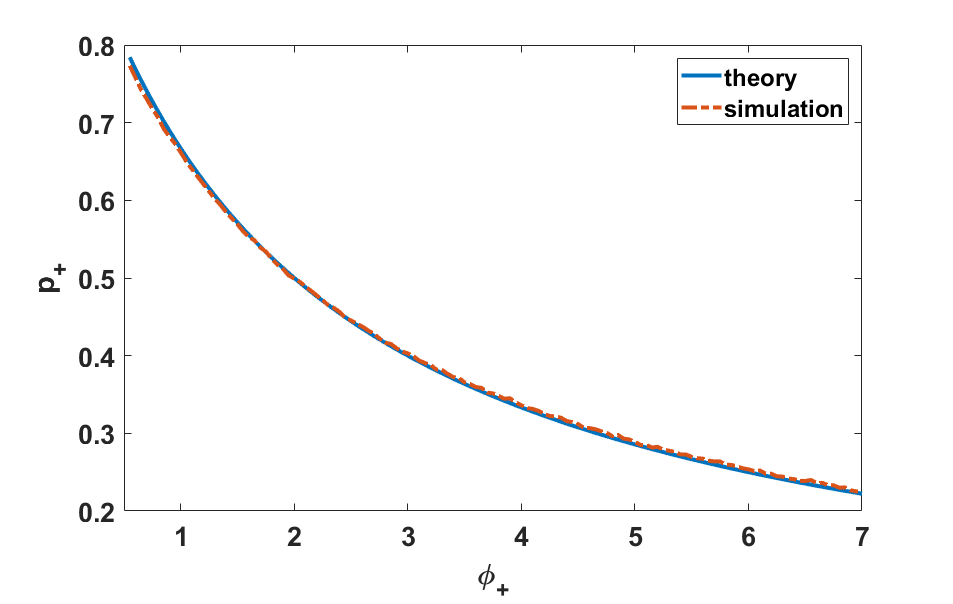}
		\includegraphics[scale=0.32]{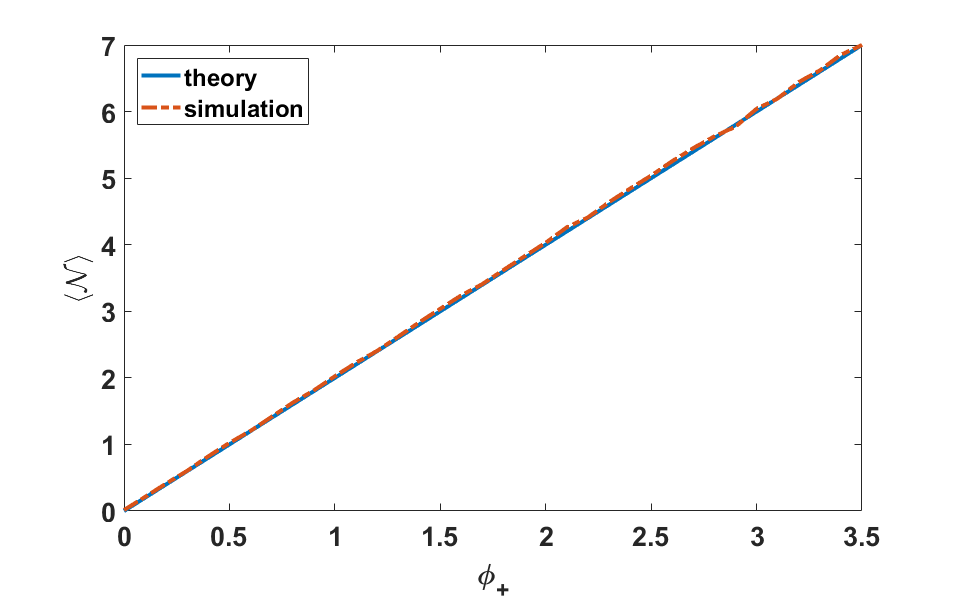}	
	\caption{ $p_+$ (left) and $\langle \cN \rangle $ (right) for the Brownian motion are presented  for  $\phi$ in units of $H/2 \pi$. 
	We have fixed $\phi_-=-2$  while varying $\phi_+$. The analytical results 
	Eqs.~(\ref{Brownian-probs}) and (\ref{N-av-Brownian}) shown by the solid blue curves are in excellent  agreement with the  full numerical results shown by the dashed red curves.  }
	\label{pplus}
\end{figure}

In Fig.~\ref{pplus}  we have presented our numerical results for $p_+$ and $\langle \cN \rangle$ and compared the numerical results with the analytical results Eqs.~(\ref{Brownian-probs}) and (\ref{N-av-Brownian}). We see that they are in excellent agreement. 

As we mentioned earlier, the results of this section may not apply to the standard observable phase of inflation.  It is a toy model to study boundary crossing probability that can have very large number of $e$-folds.  Therefore, it can belong to the regime of eternal inflation.  It is well-known that predictions in an eternally inflating universe, and even in some non-eternal situations, depend on the choice of measure \cite{Linde:2010xz}.  We do not plan to delve into the details, but let us just mention that our calculations correspond to a scale-factor cutoff measure without volume-weighting. This is because we employ the number of $e$-folds as time, and because each realization of our stochastic process occurs in a super-horizon patch.  In fact, since we have a constant Hubble, a volume-weighted measure would give the same results too.

\subsection{ Case with classical drift  }

Now consider the general case where $\dot \phi_0 \neq 0$ so we have a classical drift in addition to the noise term.  As we mentioned in section \ref{non-attractor}, when $\dot \phi_0 \neq 0$, there is a classical limit $\phi = \phi_{\mathrm {max}}$  beyond  which the field cannot go.    Intuitively speaking, we can imagine that when the field has approached the classical limit $\phi_{\mathrm {max}}$, then its classical evolution  becomes more and more negligible while the quantum diffusion terms from $W(\cN)$ becomes more relevant. 

In terms of  $\phi_{\mathrm {max}}$ given in Eq.~(\ref{phimax}),  Eq.~(\ref{phi-cN})  can be cast into
\ba
\label{chi-eq}
\frac{H}{2 \pi  \phi_{\mathrm {max}}} W(\cN) = \chi (\cN) + e^{-3 \cN} \, ,
\ea
in which we have defined the field displacement relative to $\phi_{\mathrm {max}}$ via
\ba
\chi \equiv \frac{\phi(\cN)}{\phi_{\mathrm {max}}} -1 \, .
\ea

\begin{figure}
		\centering
		\includegraphics[scale=0.32]{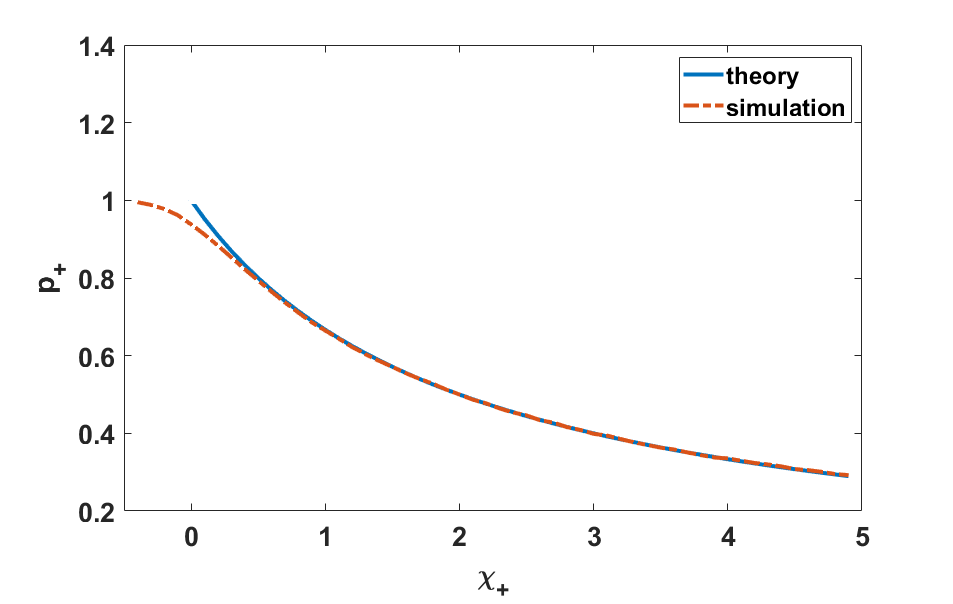}
		\hspace{0cm}
		\includegraphics[scale=0.32]{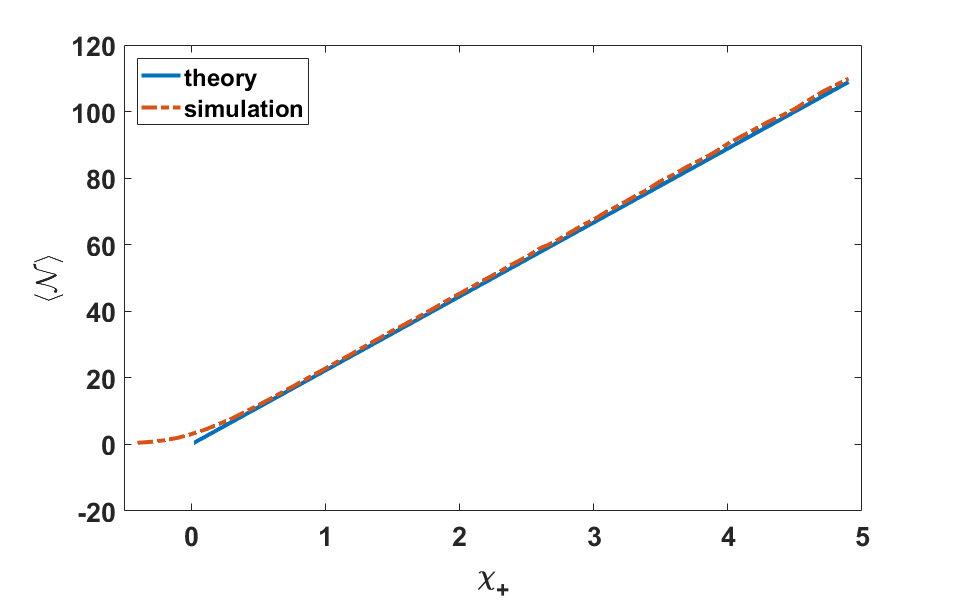}	
		\vspace{1cm}
	\caption{    $p_+$ (left) and $\langle \cN \rangle $ (right) for the model with drift, Eq.~(\ref{chi-eq}), are presented.  Similar to  Fig.~\ref{pplus} we have fixed $\chi_-=-2$ and varied $\chi_+$. The analytical results Eqs.~(\ref{p+-app}) and (\ref{N-app}) (shown by the solid blue curves) are in very good agreement with the full numerical result (shown by the dashed red curves) when $\langle \cN \rangle \gg 1$.  To reduce the cost of numerical analysis involving huge stochastic simulations, 
	we have considered a hypothetical  case with $\calP_\calR = 10^{-2}$.  }
	\label{drift-plots}
\end{figure}

Despite its simple form, we could not solve Eq.~(\ref{chi-eq}) analytically to find the first hitting probabilities $p_\pm$ 
and $\langle \cN \rangle$. The reasons are that we have a time-dependent drift term   $ e^{-3 \cN}$  and also 
that the stochastic variable $\cN$ appears in $W(\cN)$. These made it difficult to find the analytical solution in the presence of two barriers. This should be compared with the analysis in section \ref{USR-stoch} where we had effectively a single barrier, i.e., the surface of end of inflation, so we were able to find analytical results. In addition, in section \ref{USR-stoch} we could make a perturbative expansion in terms of $\kappa$ but here we can not  perform a perturbative expansion specially when the field has approached $\phi_{\mathrm {max}}$. 

However, if one could neglect the contribution of the drift term $ e^{-3 \cN}$, then Eq.~(\ref{chi-eq}) becomes a Brownian motion like Eq.~(\ref{Brownian-eq}) 
with the initial condition  $\chi=0$.  Intuitively, this corresponds to the situation that one starts at $\phi_{\mathrm {max}}$ where the velocity vanishes and the field evolves 
because of the quantum kicks as in Brownian motion. Of course, for this to happens one requires $\cN \gg1$. In this approximation, one can use Eqs.~(\ref{Brownian-probs}) and (\ref{N-av-Brownian}) (with the replacement $\phi_\pm \rightarrow \chi_\pm)$ to obtain $p_\pm$ and 
$\langle \cN \rangle$,  yielding 
\ba
\label{p+-app}
p_+ &\simeq& \frac{-\chi_-}{\chi_+ - \chi_-}  = \frac{\dot \phi_0 - 3 H \phi_-}{ 3 H ( \phi_+ - \phi_-)}
\, , \\
\label{p--app}
p_- &\simeq& \frac{\chi_+}{\chi_+ - \chi_-}  = \frac{-\dot \phi_0 + 3 H \phi_+}{ 3 H ( \phi_+ - \phi_-)} \, ,
\ea
and
\ba
\label{N-app}
\big \langle \cN  \big\rangle  &\simeq& \big( \frac{H}{2 \pi \phi_{\mathrm {max}}} \big)^{-2} \left( p_+ \chi_+^2 + p_- \chi_-^2 \right)  \nonumber\\
&=&  \frac{-\chi_- \chi_+}{9 {\calP}_\calR} \, .
\ea
Taking $\chi_+$ and $-\chi_-$ typically to be order few, we see that $\big \langle \cN  \big \rangle \sim {{\calP}_\calR}^{-1}$. This is a measure of the largeness of  $\big \langle \cN  \big \rangle$ assumed above. 

 In Fig.~\ref{drift-plots} we have presented the results for $p_+$ and $\langle \cN \rangle$ by solving Eq.~(\ref{chi-eq})
numerically. As expected, the Brownian estimations  Eqs.~(\ref{p+-app}) and 
 (\ref{N-app}) are  in very good agreement with the exact numerical results when  $\cN \gg 1$.
  This confirms that once the effect of the classical drift has died out, we can approximate the dynamics by a   Brownian motion with the initial condition set at $\chi =0$ corresponding to 
 $\phi= \phi_{\mathrm {max}}$.

\section{Summary and Discussions }
\label{summary}

In this work we have studied ultra slow-roll model in the context of stochastic inflation. 
The USR setup with a flat 
potential is an ideal place to investigate the stochastic effects during inflation. The coarse grained  
super-horizon scale perturbations receive active quantum kicks from small scales. The stochastic $\delta N$ formalism is a powerful tool to study cosmological correlations in stochastic inflation.   Using the stochastic calculus, we have calculated the mean number of $e$-folds, the power spectrum and the bispectrum in USR inflation. We have correctly reproduced the known leading classical terms in these cosmological 
correlations. In addition, we have shown that the fractional corrections to each observable correlations
induced from  stochastic dynamics   are at the order of ${\cal P}_e$, the curvature perturbation power spectrum at the end of non-attractor phase. 

There have been discussions in the literature on the  contributions of stochastic dynamics in the curvature perturbation power spectrum for primordial black hole formation during inflation 
\cite{Pattison:2017mbe,  Ezquiaga:2018gbw, Cruces:2018cvq}. Our results indicate that the 
stochastic contributions are negligible during USR phase. This is in agreement with the results of 
\cite{Cruces:2018cvq}. However, we do not agree with the conclusion in \cite{Cruces:2018cvq} that the
$\delta N$ formalism and the separate universe approach is not valid in the USR setup. On the contrary, the $\delta N$ formalism  and the separate universe approach are well applicable in the USR setup. The main requirements for the applicability of the $\delta N$ formalism is the energy conservation and the validity of the gradient expansion on super-horizon scales. These requirements are independent of whether the system has reached the attractor phase, as in conventional slow-roll models, or it is still in non-attractor phase as in current USR setup. It was demonstrated explicitly in \cite{Namjoo:2012aa, Chen:2013aj, Chen:2013eea} that the field theoretical in-in formalism and the $\delta N$ formalism yield the same results for the power spectrum and bispectrum. In addition, in the current work, using the stochastic $\delta N$ formalism,
we have correctly reproduced the results of  \cite{Namjoo:2012aa} in USR setup  while calculating the sub-leading stochastic corrections.

We also have calculated the first hitting probabilities in the USR setup containing two absorbing barriers.
An extreme case is when the field has no classical velocity so the quantum fluctuations of the test field are governed by the Brownian motion. We have shown that $p_+$ and $\langle \cN \rangle$ agree with the theoretical predictions of the Brownian motion. We extended this analysis to the case when the field has initial velocity, inducing a time-dependent drift term. Based on the physical intuition, we expect that once the classical drift term has become negligible then the system approaches the Brownian limit. We have calculated $p_+$ and $\langle \cN \rangle$ in this limit and have shown that our theoretical approximations are in very good agreement with the full numerical results when $\langle \cN \rangle \gg1$. 

There are a number of directions in which the current analysis can be extended. A natural extension is to study a more non-trivial setup of non-attractor inflation beyond the simple USR setup. As we have mentioned before,  the simple USR setup suffers from the shortcoming that inflation does not end. One requires dynamics beyond the USR setup, say a waterfall mechanism, to terminate inflation.  A more reasonable extension of the USR setup is to consider a potential which is very flat only in some finite region of field space, such as in potentials  having an inflection point, while having slow-roll slopes for other regions of the potential.   In this more physical picture, the inflaton field  rolls towards the flat region  and after some 
period of USR-like inflation, it exits from the flat region and inflation continues as in conventional slow-roll models. Of course, in this case, the stochastic dynamics and the corresponding Langevin equations become more complicated and one may not be able to solve  the system of equations analytically to find the mean number of $e$-folds, the power spectrum and bispectrum. One may need to use numerical methods to study the system. We would like to come back to this question in future. 

\vspace{1cm}
 

{\bf Acknowledgments:}  We thank S.~Baghram, T.~Fujita and A.~Starobinsky for discussions and comments. We are grateful to  H.~Assadullahi, V.~Vennin and D.~Wands for many insightful discussions during the progress of this work and beyond.  H.~F.\ and A.~N.\  thank the Yukawa Institute for Theoretical Physics at Kyoto University for hospitality during the YITP symposium YKIS2018a ``General Relativity -- The Next Generation --".  A.~N.\ thanks ICTP for hospitality during the progress of this work. H.~F.\ thanks ICG and the University of Portsmouth  for kind hospitality where this work was in its final stage. M.~N.\ acknowledges financial support from the research council of University of Tehran.

\appendix
\section{Stochastic Calculus}
\label{calculus}

In this appendix we review the basic elements of stochastic calculus and derive some results that were used in the body of the paper.  For a more detailed introductory account consult Ref.~\cite{Evans}.
Our basic equation   \eqref{Langevin} is a special case of a stochastic differential equation (also known as a Langevin equation)
\begin{equation}
\label{LangevinGeneral}
d\phi(N) = A\big(\phi(N),N \big) dN + B\big(\phi(N),N\big) dW(N),
\end{equation}
where $A=\dot\phi_0/H$ and $B=H/2\pi$.  The quantity $dW=\xi dN$ is the differential of $W(N)$, known as the Wiener process (or Brownian motion), and $\xi(N)$ is called the white noise.  The goal is to obtain the statistics of $\phi(N)$, which is a stochastic process, given some initial condition, which in our case is $\phi(0)=\phi_0$.  It is convenient to think of the realizations of the stochastic process as random trajectories represented by the random function $\phi(N)$.
 
Two of the basic properties of $dW$ and $\xi$ are
\begin{equation}
\big(dW(N)\big)^2 = dN \, , \qquad \qquad   \langle \xi(N) \xi(N') \rangle = \delta(N-N').
\end{equation}
We also have the initial condition $W(0)=0$.  

A fundamental result is Ito's lemma, assuming that Eq.~(\ref{LangevinGeneral}) follows the Ito's stochastic scheme,  that for any non-stochastic function $f$ of $\phi$ we have
\begin{equation}\label{Ito}
\begin{aligned}
f(\phi(N_2)) - f(\phi(N_1)) &= \int_{N_1}^{N_2} f'(\phi) d\phi + \frac12 \int_{N_1}^{N_2} f''(\phi) d\phi^2 \\
&= \int_{N_1}^{N_2} \big( Af'+ \frac12 B^2f'' \big) dN + \int_{N_1}^{N_2} Bf' dW.
\end{aligned}
\end{equation}
Another important result is
\begin{equation}
\label{non-anticipating}
\Big\langle \int_{N_1}^{N_2} f dW \Big\rangle = 0,
\end{equation}
which holds for any non-anticipating function $f$.  A non-anticipating function is one whose value at time $N$ doesn't depend on what happens in the future of $N$.

Finally, we define $\cal N$ as the smallest $N>0$ at which the trajectory of $\phi(N)$ crosses the surface of end of USR phase. $\cal N$ is a special case of the class of random variables that are called stopping time.  Notably, the previous two results hold for stopping times too, i.e., when $N_1=0$, and $N_2$ is replaced by $\cal N$ in Eqs.~\eqref{Ito} and  \eqref{non-anticipating}.

As an application of the above stochastic methods, we now compute $\langle W(\cN)^n \rangle$ for $n=2,\dots,6$
which we make use of in the main text.  To begin, let us apply Ito's lemma to the function $f(W)=W^n$ 
\begin{equation}
W({\cal N})^n - W(0)^n = \int_0^{\cal N} nW^{n-1} dW + \frac12 \int_0^{\cal N} n(n-1) W^{n-2} dN.
\end{equation}
Taking the expectation values, and using Eq.~(\ref{non-anticipating}),  this reduces to
\begin{equation}
\label{a.6}
\langle W({\cal N})^n \rangle = \frac{n(n-1)}2 \Big \langle \int_0^{\cal N}  W^{n-2} dN \Big\rangle.
\end{equation}
Now, we can perform an integration by parts  to obtain
\ba
\big \langle W({\cal N})^n  \big\rangle =   \frac{n(n-1)}2 \Big( \big\langle W(\cN)^{n-2} \cN  \big\rangle  - \Big \langle \int_0^{\cal N} N d W^{n-2} \Big \rangle  
\Big) \, .
\ea
Now, using the following formula \cite{{Evans}}
\ba
d \left(W(N)^m \right) = m W^{m-1} d W + \frac{m(m-1)}{2} W^{m-2} dN \, ,
\ea
and after performing integration by parts and using Eq.~(\ref{non-anticipating}) again, we obtain
\ba
\langle W({\cal N})^n \rangle  &=& \frac{n(n-1)}2  \Bigg\{  \big \langle W(\cN)^{n-2} \cN  \big \rangle  -  \frac{(n-2) (n-3)}{4}  \times\nonumber\\ 
&& \qquad \qquad  \times  \Big[  \big \langle W(\cN)^{n-4} \cN^2 \big \rangle 
 -\frac{ (n-4) (n-5)}{2} 
 \Big \langle \int_0^{\cal N} N^2  W^{n-6}  d N  \Big \rangle \Big] \Bigg\} 
\ea
This is the equation which will be used to calculate $\langle W(\cN)^n \rangle$ for $n=2,\ldots,6$. We consider each case in turn.

\begin{itemize}

\item {$n=2$}
\ba
\label{neq2}
\big \langle W(\cN)^2 \big \rangle = \langle \cN \rangle \, .
\ea

\item {$n=3$} 
\ba
\label{a.7}
\big \langle W({\cal N})^3 \big \rangle = 3 \big \langle {\cal N} W({\cal N}) \big\rangle \, .
\ea
Now, we need to calculate 
$\langle {\cal N} W({\cal N}) \rangle$. For this purpose, we can use the expansion of $\cN$ given in Eq.~(\ref{5-7}), 
obtaining
\ba
\label{W4-a}
\big \langle W({\cal N})^3 \big \rangle = -\kappa \big \langle W(\cN)^2 \big \rangle + \frac{\kappa^2}{2} \big \langle W({\cal N})^3 \big\rangle
- \frac{\kappa^3}{3}  \big\langle W({\cal N})^4 \big\rangle +... \, ,
\ea
where we have neglected higher orders of $\kappa$. Now using Eqs.~(\ref{neq2}) and (\ref{average2}) and the leading value of $\langle W({\cal N})^4 \rangle \simeq 3 N_c^2$ (which will be shown below), we obtain the final result 
\ba
\label{W3-b}
\big \langle W({\cal N})^3 \big \rangle = - \kappa N_c \big[ 1+ \kappa^2 (N_c + \frac{2}{3} )   \big] + {\cal O}(\kappa^5) \, .
\ea

\item {$n=4$} 
\ba
\label{W4-a}
\big\langle W({\cal N})^4 \big\rangle = 6 \big\langle {\cal N} W({\cal N})^2 \big\rangle  - 3 \big\langle \cN^2 \big\rangle  \, .
\ea
Now, as in above, replacing $\cN$  from Eq.~(\ref{5-7}), we obtain
\ba
\label{W4-b}
\big\langle W({\cal N})^4 \big\rangle = 3 N_c^2 +  \kappa^2 N_c  \big( 3 N_c  + \frac{5}{3}  \big) + {\cal O}(\kappa^4) \, .
\ea

\item {$n=5$} 
\ba
\label{W5-a}
\big\langle W({\cal N})^5 \big\rangle = 10 \big\langle {\cal N} W({\cal N})^3 \big\rangle  - 15 \big\langle W(\cN) \cN^2 \big\rangle  \, ,
\ea
We need $\langle W({\cal N})^5 \rangle$ to order $\kappa$ to calculate the stochastic corrections in $f_{NL}$. We obtain
\ba
\label{W5-b}
\big\langle W({\cal N})^5 \big\rangle = -10 \kappa N_c^2 + {\cal O}(\kappa^3) \, .
\ea
\item {$n=6$} 
\ba
\label{W6-a}
\big\langle W({\cal N})^6 \big\rangle = 15 \big\langle {\cal N} W({\cal N})^4 \big\rangle  - 45 \big\langle W(\cN)^2 \cN^2 \big\rangle  
+ 15  \big\langle  \cN^3 \big\rangle  \, .
\ea
To calculate the leading stochastic corrections in $f_{NL}$, we need to calculate $\langle W({\cal N})^6 \rangle$ to 
order $\kappa^0$, obtaining 
\ba
\label{W6-b}
\big\langle W({\cal N})^6 \big\rangle = 15 N_c^3 + {\cal O}(\kappa^2) \, .
\ea

\end{itemize}

\section{Stochastic non-Gaussianity}
\label{fNL-app}

In this appendix we derive  Eq.~\eqref{fnl} for the amplitude of local non-Gaussianity,  $f_{NL}$, 

The curvature perturbation with the effects of non-linearities in local shape in real space is given by 
\cite{Komatsu:2001rj, Abolhasani:2019cqw,Wands:2010af}
\ba
\label{fNL-def}
\mathcal{R}(\bfx)=\mathcal{R}_g(\bfx)+\frac{3}{5}f_{NL} 
\big(\mathcal{R}_g(\bfx)^2-\left<\mathcal{R}_g(\bfx)^2\right> \big) \, ,
\ea
where $\mathcal{R}_g$ is the Gaussian part for the field. The coefficient  $3/5$ is a historical factor, appearing because the Bardeen potential $\Phi$ is related to curvature perturbation in matter dominated era, such as  during the CMB decoupling,  via 
$\Phi = 3\calR/5$. Note that to have $\langle \calR \rangle =0$, one has to subtract $\langle {\calR}_g(\bfx)^2 \rangle$ 
from  $ {\calR}_g(\bfx)^2 $.

Starting from the $\delta N$ formalism, $\mathcal{R}=\delta\mathcal{N}$, let us calculate $\langle \calR(\bfx) ^3 \rangle$. Using Eq.~(\ref{fNL-def}), we obtain 
\ba
\label{deltaN3}
\big \langle \delta \cN^3 \big\rangle = \big\langle \calR(\bfx)^3\big \rangle  = \Big \langle  \Big[  \mathcal{R}_g(\bfx)+\frac{3}{5}f_{NL} 
\big(\mathcal{R}_g(\bfx)^2-\left<\mathcal{R}_g(\bfx)^2\right> \big) \, 
\Big]^3 \Big \rangle  \, .
\ea
Because of the Gaussian nature of ${\calR}_g$, the leading non-zero contributions in the above expression starts with 
the terms  $\langle  {\calR}_g^4 \rangle $. To calculate the leading order contributions, we have to contract one term of
${\calR}_g^2$ with two terms of ${\calR}_g$. There are 3 possibilities for these contractions, yielding 
\ba
\label{b.1}
\big\langle \calR(\bfx)^3 \big\rangle  =  \frac{9}{5}f_{NL} \Big[ \left<\mathcal{R}_g^4\right>-\left<\mathcal{R}_g^2\right>^2 \Big]
+{\cal O} (\mathcal{R}_g^6 ) \, .
\ea
On the other hand, using  Wick's theorem for the Gaussian fields, $\left<\mathcal{R}_g^4\right> $ is given by
\begin{equation}
    \label{contraction}
    \left<\mathcal{R}_g^4\right>=3\left<\mathcal{R}_g^2\right>^2.
\end{equation}
Now substituting Eq.~\eqref{contraction} in Eq.~\eqref{b.1} we obtain  
\begin{equation}
    \label{b.3}
    \left<\mathcal{R}^3\right>=\frac{18}{5}f_{NL}\left<\mathcal{R}_g^2\right>^2 \, ,
\end{equation}
where we have neglected the sub-leading ${\cal O} (\mathcal{R}_g^6)$ terms. 

On the other hand, as we discussed around  Eq.~(\ref{variance-R}),  the variance $\left<\mathcal{R}_g(\bfx)^2\right>$ is related to the accumulation of super-horizon modes via
\ba
\label{variance-Rb}
\langle \calR^{2}(\bfx)  \rangle = \int_{k_{i}  }^{k_{e}} \frac{dk }{k} 
\frac{k^{3}}{2 \pi^{2}} | \calR_{k}|^{2} \simeq  
\int_{\ln k_{e} - \langle \cN \rangle }^{\ln k_{e}} d N\, 
\calP_{\delta N} \, .
\ea
Therefore, the three point correlation function is related to the variance as follows
\ba
\big\langle \calR(\bfx)^3 \big\rangle  =  \frac{18}{5}f_{NL}   \Big[   
\int_{\ln k_{e} - \langle \cN \rangle }^{\ln k_{e}} d N\, 
\calP_{\delta N}  \Big]^2 \, .
\ea
Correspondingly, $f_{NL}$ can be obtained as
\begin{equation}
    \label{b.5}
    f_{NL}=\frac{5}{36{\calP_\calR^2}}\frac{{\mathrm d}^2\left<\delta\mathcal{N}^3\right>}{\mathrm d\left<\mathcal{N}\right>^2}.
\end{equation}
Note that in obtaining the above formula for $f_{NL}$ we have neglected the derivative of $\mathcal{P_\calR}$ since it is proportional to $n_s-1$ ($n_s$ being the spectral index) which is very small.



\begin{thebibliography}{99}

\bibitem{Akrami:2018odb} 
Y.~Akrami {\it et al.} [Planck Collaboration], 
[arXiv:1807.06211 [astro-ph.CO]].
  
\bibitem{Ade:2015lrj} 
P.~A.~R.~Ade {\it et al.} [Planck Collaboration], 
Astron.\ Astrophys.\  {\bf 594}, A20 (2016), [arXiv:1502.02114 [astro-ph.CO]].

\bibitem{Vilenkin:1983xp} 
A.~Vilenkin, 
Nucl.\ Phys.\ B {\bf 226}, 527 (1983).

\bibitem{Starobinsky:1986fx} 
  A.~A.~Starobinsky,
  Lect.\ Notes Phys.\  {\bf 246}, 107 (1986).

\bibitem{Nakao:1988yi}
K.-i. Nakao, Y.~Nambu, and M.~Sasaki, 
{ Prog.Theor.Phys.} {\bf 80} (1988) 1041.
  
\bibitem{Sasaki:1987gy} 
  M.~Sasaki, Y.~Nambu and K.~i.~Nakao,
  Nucl.\ Phys.\ B {\bf 308}, 868 (1988).

\bibitem{Nambu:1987ef}
Y.~Nambu and M.~Sasaki, 
  { Phys.Lett.} {\bf B205} (1988) 441.

\bibitem{Nambu:1988je}
Y.~Nambu and M.~Sasaki, 
  { Phys.Lett.} {\bf B219} (1989) 240.

\bibitem{Kandrup:1988sc}
H.~E. Kandrup, 
  { Phys.Rev.} {\bf D39} (1989) 2245.

\bibitem{Nambu:1989uf}
Y.~Nambu, 
{ Prog.Theor.Phys.} {\bf 81} (1989) 1037.

\bibitem{Mollerach:1990zf}
S.~Mollerach, S.~Matarrese, A.~Ortolan, and F.~Lucchin, 
{ Phys.Rev.} {\bf D44} (1991)
  1670--1679.

\bibitem{Linde:1993xx}
A.~D. Linde, D.~A. Linde, and A.~Mezhlumian, 
  { Phys.Rev.} {\bf D49} (1994)
  1783--1826, 
  gr-qc/9306035

\bibitem{Starobinsky:1994bd}
A.~A. Starobinsky and J.~Yokoyama, 
  { Phys.Rev.} {\bf D50} (1994)
  6357--6368, 
  astro-ph/9407016. 
  
\bibitem{Kunze:2006tu} 
  K.~E.~Kunze,
  JCAP {\bf 0607}, 014 (2006), 
  [astro-ph/0603575].

\bibitem{Prokopec:2007ak} 
  T.~Prokopec, N.~C.~Tsamis and R.~P.~Woodard,
  Annals Phys.\  {\bf 323}, 1324 (2008), 
  [arXiv:0707.0847 [gr-qc]].
  
\bibitem{Prokopec:2008gw} 
  T.~Prokopec, N.~C.~Tsamis and R.~P.~Woodard,
  Phys.\ Rev.\ D {\bf 78}, 043523 (2008), 
  [arXiv:0802.3673 [gr-qc]].
 
\bibitem{Tsamis:2005hd} 
  N.~C.~Tsamis and R.~P.~Woodard,
  Nucl.\ Phys.\ B {\bf 724}, 295 (2005), 
  [gr-qc/0505115].

\bibitem{Enqvist:2008kt} 
  K.~Enqvist, S.~Nurmi, D.~Podolsky and G.~I.~Rigopoulos,
  JCAP {\bf 0804}, 025 (2008), 
  [arXiv:0802.0395 [astro-ph]].
 
  
  
  

\bibitem{Finelli:2008zg}
F.~Finelli, G.~Marozzi, A.~Starobinsky, G.~Vacca, and G.~Venturi, 
  { Phys.Rev.} {\bf D79} (2009) 044007,
  arXiv:0808.1786.

\bibitem{Finelli:2010sh}
F.~Finelli, G.~Marozzi, A.~Starobinsky, G.~Vacca, and G.~Venturi, 
  { Phys.Rev.} {\bf D82} (2010) 064020,
  arXiv:1003.1327.

\bibitem{Garbrecht:2013coa}
B.~Garbrecht, G.~Rigopoulos, and Y.~Zhu, 
{ Phys.Rev.}
  {\bf D89} (2014) 063506, 
  arXiv:1310.0367.

\bibitem{Garbrecht:2014dca}
B.~Garbrecht, F.~Gautier, G.~Rigopoulos, and Y.~Zhu, 
{ Phys. Rev.} {\bf D91} (2015), no.~6 063520,
  arXiv:1412.4893.

\bibitem{Burgess:2014eoa} 
  C.~P.~Burgess, R.~Holman, G.~Tasinato and M.~Williams,
  JHEP {\bf 1503}, 090 (2015), 
  [arXiv:1408.5002 [hep-th]].

\bibitem{Burgess:2015ajz} 
  C.~P.~Burgess, R.~Holman and G.~Tasinato,
  JHEP {\bf 1601}, 153 (2016), 
  [arXiv:1512.00169 [gr-qc]].

\bibitem{Boyanovsky:2015tba} 
  D.~Boyanovsky,
  Phys.\ Rev.\ D {\bf 92}, no. 2, 023527 (2015),
  [arXiv:1506.07395 [astro-ph.CO]].
  
\bibitem{Boyanovsky:2015jen} 
  D.~Boyanovsky,
  Phys.\ Rev.\ D {\bf 93}, 043501 (2016), 
  [arXiv:1511.06649 [astro-ph.CO]].

\bibitem{Fujita:2017lfu} 
  T.~Fujita and I.~Obata,
  JCAP {\bf 1801}, no. 01, 049 (2018), 
  [arXiv:1711.11539 [astro-ph.CO]].





\bibitem{Fujita:2013cna} 
  T.~Fujita, M.~Kawasaki, Y.~Tada and T.~Takesako,
  JCAP {\bf 1312}, 036 (2013), 
  [arXiv:1308.4754 [astro-ph.CO]].

\bibitem{Fujita:2014tja} 
  T.~Fujita, M.~Kawasaki and Y.~Tada,
  JCAP {\bf 1410}, no. 10, 030 (2014), 
  [arXiv:1405.2187 [astro-ph.CO]].

\bibitem{Vennin:2015hra} 
  V.~Vennin and A.~A.~Starobinsky,
  Eur.\ Phys.\ J.\ C {\bf 75}, 413 (2015)
  [arXiv:1506.04732 [hep-th]].

\bibitem{Vennin:2016wnk} 
  V.~Vennin, H.~Assadullahi, H.~Firouzjahi, M.~Noorbala and D.~Wands,
  Phys.\ Rev.\ Lett.\  {\bf 118}, no. 3, 031301 (2017)
  [arXiv:1604.06017 [astro-ph.CO]].

\bibitem{Assadullahi:2016gkk} 
  H.~Assadullahi, H.~Firouzjahi, M.~Noorbala, V.~Vennin and D.~Wands,
  JCAP {\bf 1606}, no. 06, 043 (2016), 
  [arXiv:1604.04502 [hep-th]].
  
\bibitem{Grain:2017dqa} 
  J.~Grain and V.~Vennin,
  JCAP {\bf 1705}, no. 05, 045 (2017), 
  [arXiv:1703.00447 [gr-qc]].
  
\bibitem{Noorbala:2018zlv} 
  M.~Noorbala, V.~Vennin, H.~Assadullahi, H.~Firouzjahi and D.~Wands,
  JCAP {\bf 1809}, no. 09, 032 (2018), 
  [arXiv:1806.09634 [hep-th]].


\bibitem{Sasaki:1995aw} 
  M.~Sasaki and E.~D.~Stewart,
  Prog.\ Theor.\ Phys.\  {\bf 95}, 71 (1996), 
  [astro-ph/9507001].

\bibitem{Sasaki:1998ug} 
  M.~Sasaki and T.~Tanaka,
  Prog.\ Theor.\ Phys.\  {\bf 99}, 763 (1998), 
  [gr-qc/9801017].

\bibitem{Lyth:2004gb} 
  D.~H.~Lyth, K.~A.~Malik and M.~Sasaki,
  JCAP {\bf 0505}, 004 (2005), 
  [astro-ph/0411220].

\bibitem{Wands:2000dp} 
  D.~Wands, K.~A.~Malik, D.~H.~Lyth and A.~R.~Liddle,
  Phys.\ Rev.\ D {\bf 62}, 043527 (2000), 
  [astro-ph/0003278].

\bibitem{Lyth:2005fi} 
  D.~H.~Lyth and Y.~Rodriguez,
  Phys.\ Rev.\ Lett.\  {\bf 95}, 121302 (2005)
  [astro-ph/0504045].


\bibitem{Maldacena:2002vr} 
  J.~M.~Maldacena,
  JHEP {\bf 0305}, 013 (2003), 
  [astro-ph/0210603].



\bibitem{Abolhasani:2018gyz} 
  A.~A.~Abolhasani and M.~Sasaki,
  JCAP {\bf 1808}, no. 08, 025 (2018), 
  [arXiv:1805.11298 [astro-ph.CO]].

\bibitem{Namjoo:2012aa} 
  M.~H.~Namjoo, H.~Firouzjahi and M.~Sasaki,
  EPL {\bf 101}, no. 3, 39001 (2013), 
  [arXiv:1210.3692 [astro-ph.CO]].

\bibitem{Chen:2013aj} 
  X.~Chen, H.~Firouzjahi, M.~H.~Namjoo and M.~Sasaki,
  EPL {\bf 102}, no. 5, 59001 (2013), 
  [arXiv:1301.5699 [hep-th]].

\bibitem{Chen:2013eea} 
  X.~Chen, H.~Firouzjahi, E.~Komatsu, M.~H.~Namjoo and M.~Sasaki,
  JCAP {\bf 1312}, 039 (2013), 
  [arXiv:1308.5341 [astro-ph.CO]].

\bibitem{Kinney:2005vj} 
  W.~H.~Kinney,
  Phys.\ Rev.\ D {\bf 72}, 023515 (2005), 
  [gr-qc/0503017].
  
\bibitem{Martin:2012pe} 
  J.~Martin, H.~Motohashi and T.~Suyama,
  Phys.\ Rev.\ D {\bf 87}, no. 2, 023514 (2013), 
  [arXiv:1211.0083 [astro-ph.CO]].

\bibitem{Motohashi:2014ppa} 
  H.~Motohashi, A.~A.~Starobinsky and J.~Yokoyama,
  JCAP {\bf 1509}, 018 (2015)
  [arXiv:1411.5021 [astro-ph.CO]].
  
\bibitem{Pattison:2018bct} 
  C.~Pattison, V.~Vennin, H.~Assadullahi and D.~Wands,
  JCAP {\bf 1808}, no. 08, 048 (2018), 
  [arXiv:1806.09553 [astro-ph.CO]].


\bibitem{Bravo:2017wyw} 
  R.~Bravo, S.~Mooij, G.~A.~Palma and B.~Pradenas,
  JCAP {\bf 1805}, no. 05, 024 (2018), 
  [arXiv:1711.02680 [astro-ph.CO]].\\

\bibitem{Mooij:2015yka} 
  S.~Mooij and G.~A.~Palma,
  JCAP {\bf 1511}, no. 11, 025 (2015), 
  [arXiv:1502.03458 [astro-ph.CO]].\\

\bibitem{Akhshik:2015nfa} 
  M.~Akhshik, H.~Firouzjahi and S.~Jazayeri,
  JCAP {\bf 1507}, 048 (2015), 
  [arXiv:1501.01099 [hep-th]].
  
\bibitem{Akhshik:2015rwa} 
  M.~Akhshik, H.~Firouzjahi and S.~Jazayeri,
  JCAP {\bf 1512}, no. 12, 027 (2015), 
  [arXiv:1508.03293 [hep-th]].

\bibitem{Finelli:2017fml} 
  B.~Finelli, G.~Goon, E.~Pajer and L.~Santoni,
  Phys.\ Rev.\ D {\bf 97}, no. 6, 063531 (2018), 
  [arXiv:1711.03737 [hep-th]].

\bibitem{Cai:2016ngx} 
  Y.~F.~Cai, J.~O.~Gong, D.~G.~Wang and Z.~Wang,
  JCAP {\bf 1610}, no. 10, 017 (2016), 
  [arXiv:1607.07872 [astro-ph.CO]].

\bibitem{Cai:2017bxr} 
  Y.~F.~Cai, X.~Chen, M.~H.~Namjoo, M.~Sasaki, D.~G.~Wang and Z.~Wang,
  JCAP {\bf 1805}, no. 05, 012 (2018), 
  [arXiv:1712.09998 [astro-ph.CO]].







\bibitem{Biagetti:2018pjj} 
  M.~Biagetti, G.~Franciolini, A.~Kehagias and A.~Riotto,
  JCAP {\bf 1807}, no. 07, 032 (2018), 
  [arXiv:1804.07124 [astro-ph.CO]].

\bibitem{Ezquiaga:2018gbw} 
  J.~M.~Ezquiaga and J.~Garcia-Bellido,
  JCAP {\bf 1808}, 018 (2018), 
  [arXiv:1805.06731 [astro-ph.CO]].
  
\bibitem{Pattison:2017mbe} 
  C.~Pattison, V.~Vennin, H.~Assadullahi and D.~Wands,
  JCAP {\bf 1710}, no. 10, 046 (2017), 
  [arXiv:1707.00537 [hep-th]].

\bibitem{Evans}
L.~Evans, ``An introduction to stochastic differential equations,'' American Mathematical Society (2013).

\bibitem{Linde:2010xz} 
  A.~Linde and M.~Noorbala,
  JCAP {\bf 1009}, 008 (2010)
  [arXiv:1006.2170 [hep-th]].

\bibitem{Cruces:2018cvq} 
  D.~Cruces, C.~Germani and T.~Prokopec,
  arXiv:1807.09057 [gr-qc].

\bibitem{Komatsu:2001rj} 
  E.~Komatsu and D.~N.~Spergel,
  Phys.\ Rev.\ D {\bf 63}, 063002 (2001), 
  [astro-ph/0005036].

\bibitem{Abolhasani:2019cqw} 
  A.~A.~Abolhasani, H.~Firouzjahi, A.~Naruko and M.~Sasaki,
  doi:10.1142/10953


\bibitem{Wands:2010af} 
  D.~Wands,
  Class.\ Quant.\ Grav.\  {\bf 27}, 124002 (2010), 
  [arXiv:1004.0818 [astro-ph.CO]].


\end{thebibliography}
\end{document}